\documentclass[12pt,]{article}
\usepackage{lmodern}
\usepackage{setspace}
\usepackage{amssymb,amsmath}
\usepackage{ifxetex,ifluatex}
\usepackage{fixltx2e} 
\ifnum 0\ifxetex 1\fi\ifluatex 1\fi=0 
  \usepackage[T1]{fontenc}
  \usepackage[utf8]{inputenc}
\else 
  \ifxetex
    \usepackage{mathspec}
    \usepackage{xltxtra,xunicode}
  \else
    \usepackage{fontspec}
  \fi
  \defaultfontfeatures{Mapping=tex-text,Scale=MatchLowercase}
  
\fi
\IfFileExists{upquote.sty}{\usepackage{upquote}}{}
\IfFileExists{microtype.sty}{%
\usepackage{microtype}
\UseMicrotypeSet[protrusion]{basicmath} 
}{}
\usepackage[margin=1in]{geometry}
\ifxetex
  \usepackage[setpagesize=false, 
              unicode=false, 
              xetex]{hyperref}
\else
  \usepackage[unicode=true]{hyperref}
\fi
\hypersetup{breaklinks=true,
            bookmarks=true,
            pdfauthor={},
            pdftitle={Do Hospital Data Breaches Reduce Patient Care Quality?},
            colorlinks=true,
            citecolor=blue,
            urlcolor=blue,
            linkcolor=magenta,
            pdfborder={0 0 0}}
\usepackage{longtable,booktabs}
\usepackage{graphicx,grffile}
\makeatletter
\def\maxwidth{\ifdim\Gin@nat@width>\linewidth\linewidth\else\Gin@nat@width\fi}
\def\maxheight{\ifdim\Gin@nat@height>\textheight\textheight\else\Gin@nat@height\fi}
\makeatother
\setkeys{Gin}{width=\maxwidth,height=\maxheight,keepaspectratio}
\let\rmarkdownfootnote\footnote%
\def\footnote{\protect\rmarkdownfootnote}

\usepackage{titling}

  \title{Do Hospital Data Breaches Reduce Patient Care Quality?}
  \pretitle{\vspace{\droptitle}\centering\huge}
  \posttitle{\par}
  \author{}
  \preauthor{}\postauthor{}
  \date{}
  \predate{}\postdate{}

\ifx\paragraph\undefined\else
\let\oldparagraph\paragraph
\renewcommand{\paragraph}[1]{\oldparagraph{#1}\mbox{}}
\fi
\ifx\subparagraph\undefined\else
\let\oldsubparagraph\subparagraph
\renewcommand{\subparagraph}[1]{\oldsubparagraph{#1}\mbox{}}
\fi

\usepackage{color} \usepackage{multirow} \usepackage{relsize}
\usepackage{longtable}

\begin{document}
\maketitle

\begin{center}
\begin{singlespace}
 {\Large Sung Choi} \linebreak
 {\Large M. Eric Johnson} \linebreak
 \vspace{5ex} \linebreak
 Vanderbilt University \linebreak
 Owen Graduate School of Management \linebreak
 401 21st Ave S  \linebreak
 Nashville, TN 37203 \linebreak
 May 17, 2017
\end{singlespace}
\end{center}

word count: 3991

\textbf{Acknowledgement}

\begin{singlespace}
This research article was prepared by the authors' personal capacity and was partially supported by a collaborative award from National Science Foundation award CNS-1329686. The views and conclusions in this document are those of the authors and should not be interpreted as necessarily representing the official policies, either expressed or implied, of the National Science Foundation. 
\end{singlespace}

\clearpage

\textbf{Abstract}

\begin{singlespace}
Objective: To estimate the relationship between a hospital data breach and hospital quality outcome

Materials and Methods: Hospital data breaches reported to the U.S. Department of Health and Human Services breach portal and the Privacy Rights Clearinghouse database were merged with the Medicare Hospital Compare data to assemble a panel of non-federal acute-care inpatient hospitals for years 2011 to 2015. The study panel included 2,619 hospitals. Changes in 30-day AMI mortality rate following a hospital data breach were estimated using a multivariate regression model based on a difference-in-differences approach.

Results: A data breach was associated with a 0.338[95\% CI, 0.101-0.576] percentage point increase in the 30-day AMI mortality rate in the year following the breach and a 0.446[95\% CI, 0.164-0.729] percentage point increase two years after the breach. For comparison, the median 30-day AMI mortality rate has been decreasing about 0.4 percentage points annually since 2011 due to progress in care. The magnitude of the breach impact on hospitals' AMI mortality rates was comparable to a year's worth historical progress in reducing AMI mortality rates.

Conclusion: Hospital data breaches significantly increased the 30-day mortality rate for AMI. Data breaches may disrupt the processes of care that rely on health information technology. Financial costs to repair a breach may also divert resources away from patient care. Thus breached hospitals should carefully focus investments in security procedures, processes, and health information technology that jointly lead to better data security and improved patient outcomes.
\end{singlespace}

\clearpage

\section{Introduction}\label{introduction}

Health data breaches in recent years have raised serious concerns about
the security of protected health information. Wide adoption of
electronic health record systems (EHRs) following the Health Information
Technology for Economic and Clinical Health (HITECH) Act of 2009 has
digitized vast stores of patient data, increasing the possibility of
large-scale hacks and inadvertent losses. Demand for health data in the
black market makes hospitals a lucrative target for external
attackers.{[}1,2{]} Internal vulnerabilities in hospital information
systems may be exploited by the external attackers or insiders who may
inappropriately disclose data. Health data breaches include loss, theft,
unauthorized access, and hacking incidents, which may be associated with
error or negligence by hospital staff handling the data. Hospitals
reported 264 data breaches between 2011 and 2015, exposing personal
information of 5,856,093 individuals (counts based on HHS data analyzed
by authors).

Breaches arise from many different sources, but regardless of the source
the resulting discovery and mitigation of a breach can be viewed as a
random shock to a hospital's care-delivery system. While agents
affiliated with a hospital may benefit from intentionally leaking
information (e.g.~hospital staff selling patient data to a third party
for personal gains), agents (and the hospital itself) face criminal
indictment, fines, and business losses from intentional or negligent
breaches,{[}3{]} which disincentivize intentional breaches. Thus a
hospital data breach can be framed as a natural experiment to estimate
the impact of a breach on patient outcomes.

Hospital data breaches provide a unique opportunity to study how
information problems affect patient outcomes. Subsequent to a breach,
organizations typically take action to mitigate the failure and improve
security. Such actions can be far ranging, from adopting new policies
and procedures to installing new security technologies. Taking advantage
of financial incentives provided by of HITECH, many hospitals have made
investments in more secure health information technology (HIT),
replacing or enhancing modules of their EHRs. New systems often support
security features such as stronger authentication procedures and
time-outs for inactivity. Data handling policies typically change as
well to control which users have the privilege to see particular
patients or specific data fields. All of these actions require hospital
staff to acclimate to new systems, learn new procedures, and adjust to
new ways of obtaining and manipulating patient data.

In extreme cases, hospital data breaches can also negatively affect the
accuracy and timeliness of patient information available to providers. A
hacking incident may temporarily disrupt hospital's servers, making
patient data unavailable to providers while the servers are being fixed.
Severe hacking attacks may force hospitals to revert to paper
charts.{[}4,5{]}. Instances of unauthorized access suggests that the
systems in place may have weaknesses in verifying the identity of the
provider or the patient, which may increase the risk of a provider
inadvertently accessing or editing the wrong patient's information.
Inaccuracies or delays in patient information are likely to disrupt the
care process and adversely affect patient outcomes.

Information problems can ripple through the continuum of care when a
patient is transferred from one provider to another, in a
\emph{handoff}.{[}6{]} The literature on discontinuity of care has
documented the association between miscommunication during handoffs and
adverse patient events.{[}7--9{]} A hospital data breach and subsequent
mitigation activities may disrupt the information flow or the processes
involved in handoffs, therefore leading to adverse patient outcomes.

Furthermore, efforts to fix the damages from a data breach may divert
resources and attention away from initiatives focused on patient care,
like improved patient safety or medication adherence. Breached hospitals
incur significant costs associated with investigating and fixing the
breach. The Ponemon Institute estimated that in 2016 the health care
industry spent an average of \$355 per stolen record for direct and
indirect costs associated with a breach.{[}10{]} Breach costs vary
depending on the size and type of the breach. Large breaches place a
larger financial burden on the organization. Typical data breaches of
500 patients cost hospitals about \$200,000 on average, including costs
from investigating the breach, notifying the affected individuals,
public relations, credit monitoring, litigation, and fines.{[}11{]}
Large breaches cost much more -- an analysis of press releases by the
HHS since 2008 documented 40 settled cases with the median settlement
amount of \$857,750 for the fine alone.{[}12{]} The aim of our paper is
to estimate the impact of a data breach on the 30-day mortality rate
using a difference-in-differences approach to analyze a panel of
non-federal acute-care inpatient hospitals from 2011-2015. We
hypothesized that remedial activities following a breach may delay and
disrupt the patient care. There may be an initial breach effect: hacking
incidents can shutdown IT systems and disrupt care, which we see in
ransomware attacks, but these were rare before to 2016. Theft,
unauthorized access by insiders, lost devices are unlikely to have a
direct effect. However, new access and authentication procedures, new
protocols, new software after any breach incident is likely to disrupt
clinicians.

\section{Background}\label{background}

\subsection{Breach Regulation}\label{breach-regulation}

As part of the HITECH Act, health care providers, health plans, and
other entities covered by the Health Insurance Portability and
Accountability Act of 1996 (HIPAA) to are required to notify the
affected individuals, the U.S. Department of Health and Human Services
(HHS), and sometimes the media following a breach of unsecured protected
health information.{[}13{]} HHS maintains a public database of the
reported breaches affecting 500 or more individuals, submitted from
October 2009 to the present.{[}14{]}

HHS defines a data breach as impermissible use or disclosure of
protected health information.{[}13{]} HHS classifies data breaches into
the following categories: theft, loss, unauthorized access/disclosure,
improper disposal, hacking/IT incident, and unknown/other. A specific
definition of each category is not given on the HHS website and
questions can be raised whether these categories can be conceptually
isolated and subjectively defined. The boundary between theft and loss
may be unclear when a laptop goes missing. In a hospital, monitoring and
enforcing which provider is authorized to access which patient
information at what time is not a trivial task. Peeking is a widespread
problem, which if detected is now considered a breach. Hacking is a
broad term for computer intrusion by an outside party and it can be done
by various techniques. To add to the complexity, these categories are
not mutually exclusive as a breach event involves a combination of these
categories. Consequently, the number of individuals affected by the
breach is often an estimate and precise numbers should be viewed with
skepticism. An individual may be double counted if affected by a hacking
incident that attacks multiple systems.

Collections of reported breaches, like those on the HHS website and
Privacy Rights Clearinghouse, do not represent all breaches. To end up
on these public databases, a breach must be (1) detected by the entity
and (2) disclosed in some way. As defined by HHS, breaches must be
publicly reported if they affect 500 or more individuals. However, HHS
allows exemptions if the breached data satisfy the HIPAA safe harbor
method for encryption.{[}15{]} The Privacy Rights Clearinghouse reports
breaches of any size disclosed by the organization or the media.

\section{Methods}\label{methods}

\subsection{Data}\label{data}

Our analysis included breaches reported to the HHS and the Privacy
Rights Clearinghouse (PRC) database between 2011 and 2015. Similar to
the HHS database, the Privacy Rights Clearinghouse (PRC) aggregates
reported breaches from public sources including the media, blogs, and
government.{[}16{]} Both data sources provide information on the name of
the breached entity, when the breach report was filed, location of the
breached entity, type of breached entity, type of breach, and specific
comments regarding the breach. The databases only provide the name of
the breached entity; they do not provide standardized identifiers to
facilitate linkages with other data. To overcome these limitations,
observations in the breach databases were linked by hospital name and
state, however the potential for erroneous matches remains.

Breaches were more frequently reported from states with larger
populations. Due to state variations in breach notification laws and how
they are enforced, some states may be over represented. Before HITECH,
California was one of the first states to enact a breach notification
law, which became effective on July 1, 2003, and most states have
followed its language.{[}17{]} Yet as of January 4, 2016, Alabama, New
Mexico, and South Dakota do not have a breach notification law.{[}18{]}

The Centers for Medicare and Medicaid Services (CMS) provide public use
data on Medicare-certified hospitals. Healthcare Cost Report Information
System (HCRIS) provides data on hospital characteristics and financial
variables.{[}19{]} Medicare Hospital Compare provides data on hospital
quality measures.{[}20{]} Data on hospital breaches from HHS and PRC
databases were merged with HCRIS and Hospital Compare data for years
2011-2015.

The raw data panel consisted of 6,435 hospitals with 30,384
hospital-year observations. Data were restricted to non-federal
acute-care inpatient hospitals. Hospitals in the U.S territories and
Maryland (which has a prospective payment system waiver) were excluded
for consistency. To maintain consistency in the financial data, the data
were further restricted to hospitals that filed HCRIS with between 360
and 370 reporting days. When a hospital submitted multiple reports in a
given year, the most recent report was used. The restrictions yielded
3,369 acute-care hospitals with 15,517 observations. Finally,
observations with missing values in the dependent or independent
variables were dropped from analysis. 3,932 observations were missing
the 30-day AMI mortality rate, accounting for most of the missing
values. The final study panel consisted of 2,619 hospitals with 11,568
hospital-year observations.

\subsection{Generalized Difference in Difference
Model}\label{generalized-difference-in-difference-model}

The association between breaches and hospital outcomes was estimated
using a generalized difference-in-differences (DID) framework with
multiple pre- and post- periods.{[}21{]} Data breaches represent random
shocks reported in a specific year, though susceptible to measurement
error from the actual year of breach. Panel data provide pre- and
post-breach measures of patient outcomes. The DID strategy controls for
time trends in outcomes among the breached hospitals, assuming that the
breached hospitals would have followed the same trend if they had not
been breached, to isolate the change in outcomes associated with the
breach.

\[ Y_{it} = \alpha_i + year_t + \beta X_{it} + \sum\limits_{n=-4}^{-2} \pi_n D_i (t-T_i^*=n )_{it} + \sum\limits_{n=0}^{4} \tau_n D_i (t-T_i^*=n)_{it} + \epsilon_{it} \]

The DID model was specified as the following equation. For hospital
\emph{i} at time \emph{t=2011\ldots{}2015}, \(Y\) is the 30-day
mortality rate (\%) in Medicare Hospital Compare,{[}22{]} adjusted for
patient characteristics to allow comparisons between hospitals. The risk
adjusted mortality rates are model based estimates and they have
uncertainty around them, which we ignore in our analysis. The mortality
rates are based on a 36-month moving average, starting from the current
year and moving back 36 months; hence a lagged response to a breach is
expected. This is a limitation set by the data provider. Hospital
Compare data reports mortality and readmission rates for AMI, heart
failure, and pneumonia.

AMI, pneumonia, and heart failure are common conditions. Past studies
have found that the adoption of health IT has been associated with
improvement in some of the quality measures for these conditions. For
example, health IT adoption was found to reduce the 60-day mortality for
pneumonia and congestive heart failure.{[}23{]} Thus, such improvements
in pneumonia and heart failure outcomes associated with newly adopted
HIT may offset the negative impact of a data breach. AMI is an acute
condition that requires timely treatment using accepted guidelines, thus
clinicians treating AMI patients may see less benefit from improvements
in HIT. If adopting HIT offers little improvement to AMI mortality, such
null relationship helps our estimates to isolate the negative impact of
a data breach on AMI mortality due to subsequent changes in HIT and
patient care processes. We conjectured that AMI mortality would be the
most sensitive to data breach among the three conditions. Preliminary
analysis showed that breaches were correlated with the AMI mortality
rate. AMI is an acute event, in which a hospitalized patient's outcome
depends on the quality of inpatient care.{[}24{]} Also, acute medical
events are less prone to selection bias due to patient choice, which
reduces the possibility of patients avoiding service at a hospital known
to have poor quality or one that had been breached. Hospital data
breaches were associated with lower number of outpatient visits and
admissions in the long-run, which suggests that patients may avoid
hospitals involved in a data breach{[}25{]}. Thus we chose AMI as our
focus for the DID analysis.

\(D_i\) is a treatment dummy, which is set to 1 if hospital was
breached. If a hospital was breached in multiple years, only the first
year was coded as breached and the subsequent years were coded as not
breached. Thirty eight of the 2,619 hospitals were breached in multiple
years. This specification assumes that a breach is a one-time event,
which is true for most but not all hospitals in the data. For
simplicity, this specification ignores multiple breaches, which may be
correlated with the severity of the information problem.

\((t-T_i^*=n )\) are time-to-event dummies, which are set to 1 when the
year of observation \(t\) is \(n=-4,-3, -2,0,...4\) years away relative
to the hospital specific time of breach \(T_i^*\). \(n=0\) is the year
of the breach. The year before breach \(n=-1\) was set as the omitted
category. The effect of the breach \(n\) years after the event is
captured by the coefficient \(\tau_n\) on the interaction of the
treatment dummy with the time-to-event dummies. A simpler model that
only estimates a single event dummy raises the concern of how to specify
the not-breached years. Categorizing all of the periods before and after
a breach year into a single not-breached category is an
oversimplification when rich panel data is available. Categorizing the
time periods into a dichotomous pre and post breach is problematic
because the impact of a breach is unlikely to be permanent. A set of
time-to-event dummies can flexibly capture how the breach impact varied
over time.

Setting \(n=-1\) as the omitted category has the advantage of larger
observations in that category because there are fewer observations at
the far ends of the relative time. For example, only the hospitals
breached at 2015 have an observation at -4. Hence the point estimates at
the far ends are imprecise due to smaller sample size, and using them as
the reference category will make it harder to detect significant
differences.

We modified the specification from Jacobson et al{[}21{]} because our
treatment-event-time is different for every hospital. Also, observations
in the control group do not have a time-to-event dummy; instead they
were coded as a time-invariant ``never breached'' dummy, which was
omitted from the fixed effects estimation. This specification is
equivalent to combining the never-breached observations with the omitted
\(n=-1\) category instead of giving them their own dummy.

Coefficient \(\pi_n\) tests the pre-breach trend between the breached
hospitals and the never breached hospitals. Prior to the event, the
breached hospitals and the control hospital are expected to have
parallel trends thus \(\pi_n\) should not significantly diverge from 0.
We assumed the effect of a breach (at n years since the event) is the
same regardless of what year the breach occurred.

\(\alpha_i\) is the hospital fixed effects. \(year_t\) is the year fixed
effects. An organization's \emph{safety culture} captures the knowledge,
beliefs, and attitudes regarding safety in the organization.{[}26{]}
Safety and security are rooted in cultures that emphasize the importance
of well-designed processes and heightened awareness of goals. We suggest
that patient safety and data security cultures are closely related. The
overall hospital safety climate, influenced by organizational policy
regarding safety, has been associated with readmissions for AMI and
heart failure.{[}24,27{]} Hospital fixed effects control for the
unobserved time-invariant hospital safety culture.

The DID models was estimated using a fixed effects regression where the
hospital effect is removed using the within transformation. The within
transformation also removes the time-invariant regressors, such as the
treatment dummy \(D\), and they cannot be separately identified.

\(X_{it}\) are the time varying hospital characteristics, including
operating revenue, number of beds, length of stay, bed occupancy rate,
\emph{meaningful use} status (meaningful user of electronic health
records defined in HITECH), patient satisfaction, and patient safety
indicators. Patient satisfaction measures included the Hospital Consumer
Assessment of Healthcare Providers and Systems (HCAHPS) survey items
from the Medicare compare data.{[}28{]}

Hospital data breaches may be correlated with underlying care quality
problems that have negative implications for patient outcomes. Provider
error or negligence in securing patient data may be associated with the
error or negligence in providing care. Hospitals with poor Agency for
Healthcare Research and Quality (AHRQ) patient safety indicators have
been associated with higher readmission rates and mortality.{[}29,30{]}
A subset of hospitals reported the AHRQ patient safety indicators
PSI-4,6,12,15,90, which were used as control variables in robustness
tests. Standard errors are heteroskedasticity robust and allow for
within hospital correlation.

\section{Results}\label{results}

\subsection{Descriptive statistics}\label{descriptive-statistics}

\begin{figure}[htbp]
\centering
\includegraphics{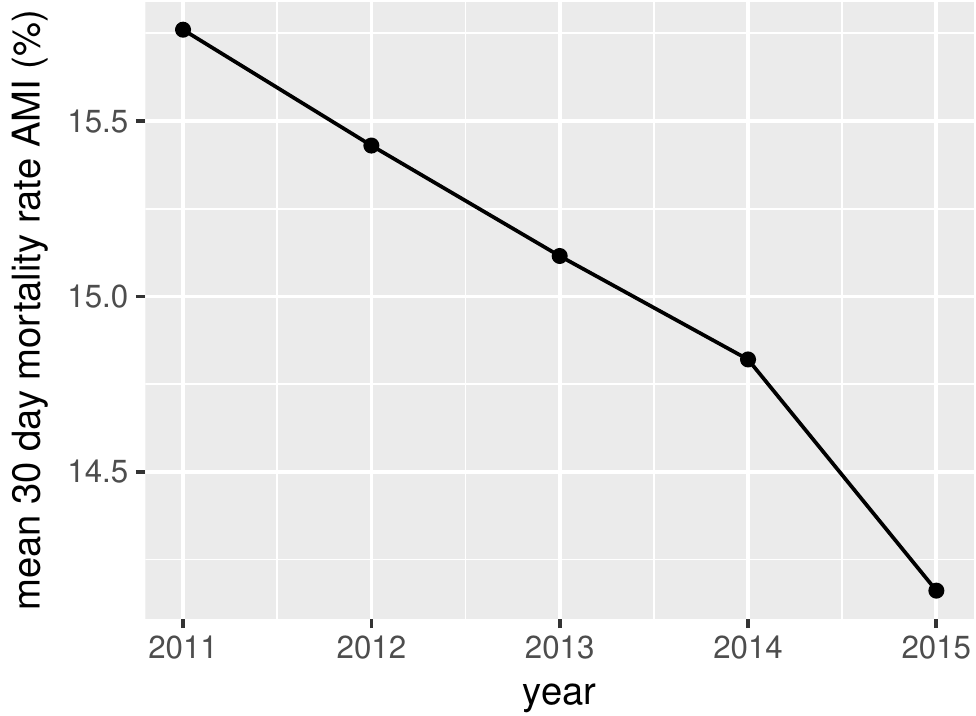}
\caption{Year vs mean 30-day AMI mortality
rate\label{fig:yr_vs_mort_plot}}
\end{figure}

Figure 1 shows that the 30-day AMI mortality rate has been decreasing,
for all hospital-year observations. The mean 30-day AMI mortality rate
decreased from 15.76\% in 2011 to 14.16\% in 2015, showing steady
improved treatment for AMI.

\begin{figure}[htbp]
\centering
\includegraphics{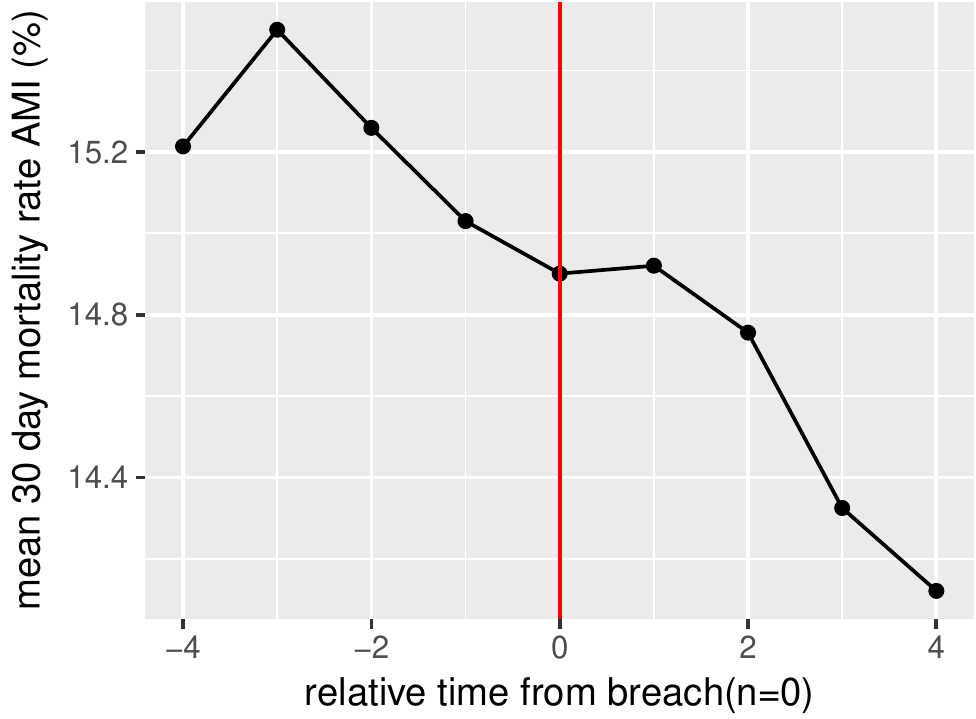}
\caption{Relative time from breach(n=0) vs mean 30-day AMI mortality
rate for the subset of hospitals that had been
breached\label{fig:breachdiff_vs_mort_plot}}
\end{figure}

Figure 2 shows the 30-day AMI mortality rate over the time from breach,
for the breached hospital-year observations. The 30-day AMI mortality
rate is decreasing, but the flat trend around the year of the breach
suggests a change in slope associated with the breach.

In figure 3, never-breached hospitals were assigned to a randomly
selected breach year, then each hospital-year observation was assigned a
pseudo-relative time. The mean 30-day AMI mortality rate of the breached
and non-breached hospital-year observations were plotted on the same
relative-time x-axis. The breached and non-breached followed a similar
decreasing trend, except for the higher mortality rate among the
breached hospital-year observations during the first and second years
after a breach.

\begin{figure}[htbp]
\centering
\includegraphics{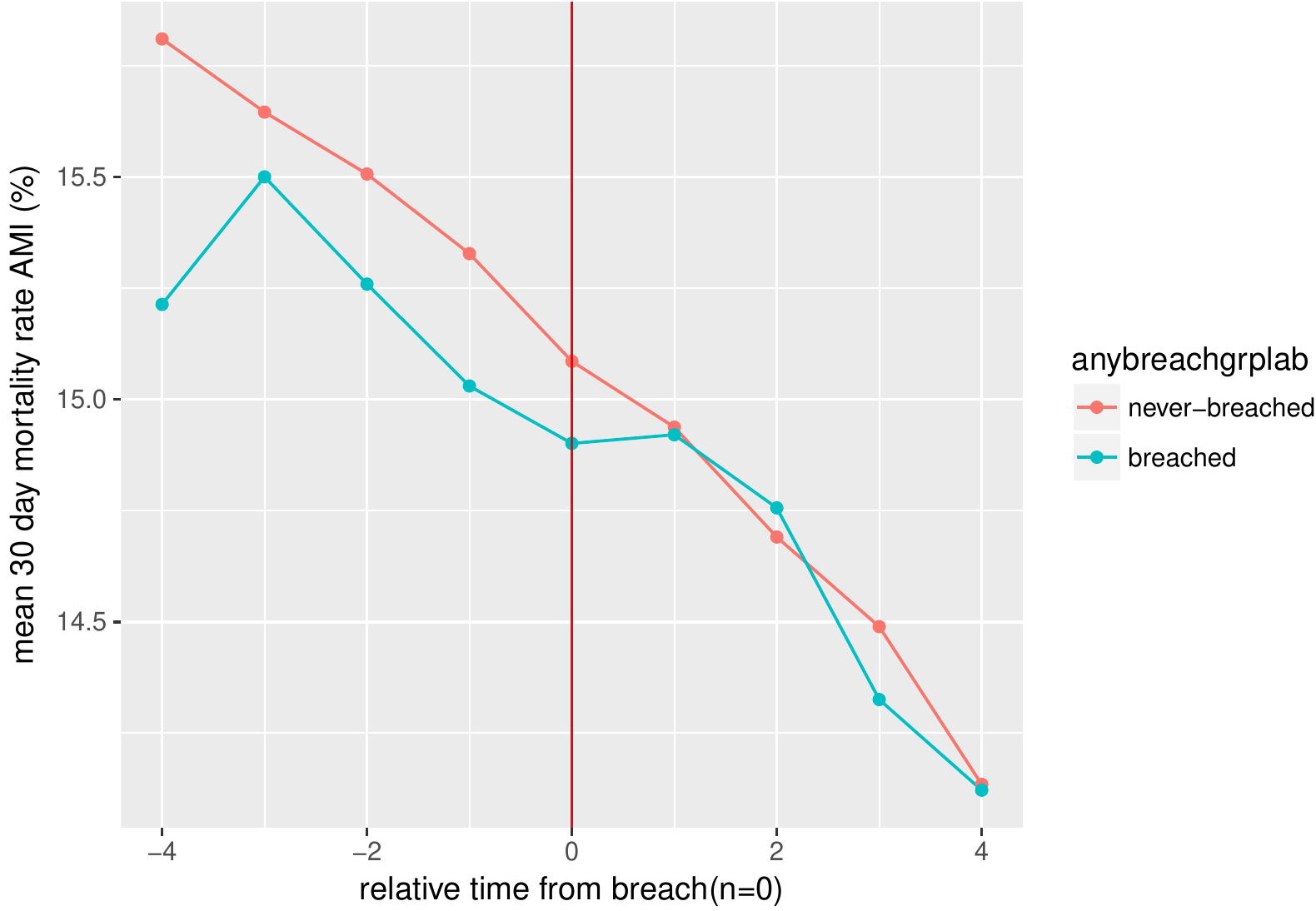}
\caption{Relative time from breach(n=0) vs mean 30-day AMI mortality
rate with never-breached hospitals assigned to a randomly selected
breach year}
\end{figure}

\clearpage

\begin{longtable}[c]{@{}lrr@{}}
\caption{Count of reported breaches by type 2011-2015}\tabularnewline
\toprule
Breach Type & N & \%\tabularnewline
\midrule
\endfirsthead
\toprule
Breach Type & N & \%\tabularnewline
\midrule
\endhead
Hacking IT Incident & 23 & 8.7\%\tabularnewline
Improper Disposal & 4 & 1.5\%\tabularnewline
Loss & 73 & 27.7\%\tabularnewline
Multiple Types & 3 & 1.1\%\tabularnewline
Other & 13 & 4.9\%\tabularnewline
Theft & 62 & 23.5\%\tabularnewline
Unauthorized Access Disclosure & 86 & 32.6\%\tabularnewline
Sum & 264 & 100\%\tabularnewline
\bottomrule
\end{longtable}

\begin{longtable}[c]{@{}lrr@{}}
\caption{Sum of individuals affected by breach type
2011-2015}\tabularnewline
\toprule
Breach Type & N & Individuals Affected\tabularnewline
\midrule
\endfirsthead
\toprule
Breach Type & N & Individuals Affected\tabularnewline
\midrule
\endhead
Hacking IT Incident & 12 & 4,922,533\tabularnewline
Improper Disposal & 4 & 3,192\tabularnewline
Loss & 18 & 86,070\tabularnewline
Multiple Types & 3 & 49,644\tabularnewline
Other & 10 & 88,293\tabularnewline
Theft & 62 & 354,719\tabularnewline
Unauthorized Access Disclosure & 39 & 351,642\tabularnewline
Sum & 148 & 5,856,093\tabularnewline
\bottomrule
\end{longtable}

Table 1 shows the hospital-year counts by breach type. A total of 264
hospital-years were breached. The three most common breach types were
unauthorized access (86), loss (73), and theft (62). A subset of
breached observations reported the number of individual records affected
by the breach. While this measure is prone to error and underreporting,
it is a proxy for the severity of a breach. Table 2 shows the sum of
individuals affected by breach type. 148 hospital-years reported the
number of individuals affected by breaches, summing to approximately
5,856,093 individual records. Between 2011-2015, 12 hacking IT incidents
affected approximately 4,922,533 individuals.

\clearpage

\begin{longtable}[c]{@{}llll@{}}
\caption{Hospital-year characteristics}\tabularnewline
\toprule
& never breached & pre-breach & post-breach\tabularnewline
\midrule
\endfirsthead
\toprule
& never breached & pre-breach & post-breach\tabularnewline
\midrule
\endhead
n & 10511 & 366 & 691\tabularnewline
30-day mortality rate AMI, mean (sd) & 15.1 (1.5) & 15.2 (1.7) & 14.7
(1.6)\tabularnewline
Operating revenue, mean (sd), \$mn & 241.7 (254.6) & 574.7 (527.8) &
686.0 (659.5)\tabularnewline
Number of beds, mean (sd) & 250.4 (183.8) & 469.7 (321.7) & 515.6
(420.2)\tabularnewline
Length of stay, mean (sd) & 4.4 (0.8) & 4.9 (0.9) & 4.9
(0.9)\tabularnewline
Bed occupancy rate, mean (sd) & 56.5 (16.6) & 68.4 (15.0) & 67.4
(14.5)\tabularnewline
Meaningful user this year, n (\%): yes & 6727 (64.0) & 150 (41.0) & 524
(75.8)\tabularnewline
Ownership, mean (sd): & & &\tabularnewline
Non-profit & 6876 (65.4) & 237 (64.8) & 472 (68.3)\tabularnewline
Profit & 2395 (22.8) & 22 ( 6.0) & 76 (11.0)\tabularnewline
Public & 1240 (11.8) & 107 (29.2) & 143 (20.7)\tabularnewline
Teaching status, n (\%): & & &\tabularnewline
Major teaching & 921 ( 8.8) & 148 (40.4) & 266 (38.5)\tabularnewline
Minor teaching & 2838 (27.0) & 107 (29.2) & 219 (31.7)\tabularnewline
Non-teaching & 6752 (64.2) & 111 (30.3) & 206 (29.8)\tabularnewline
Year, n (\%): & & &\tabularnewline
2011 & 2175 (20.7) & 159 (43.4) & 57 ( 8.2)\tabularnewline
2012 & 2186 (20.8) & 116 (31.7) & 100 (14.5)\tabularnewline
2013 & 2125 (20.2) & 62 (16.9) & 150 (21.7)\tabularnewline
2014 & 2073 (19.7) & 29 ( 7.9) & 181 (26.2)\tabularnewline
2015 & 1952 (18.6) & 0 ( 0.0) & 203 (29.4)\tabularnewline
Percent definitely not recommend, mean (sd) & 5.5 (3.0) & 5.6 (3.1) &
5.1 (2.7)\tabularnewline
Percent probably recommend, mean (sd) & 25.0 (6.6) & 24.2 (7.0) & 22.4
(6.2)\tabularnewline
Percent definitely recommend, mean (sd) & 69.5 (8.7) & 70.2 (9.4) & 72.5
(8.2)\tabularnewline
Percent patient rating \textless{}= 6, mean (sd) & 8.9 (3.8) & 9.7 (4.3)
& 8.5 (3.3)\tabularnewline
Percent patient rating 7-8, mean (sd) & 23.3 (4.6) & 24.0 (5.1) & 22.1
(4.4)\tabularnewline
Percent patient rating 9-10, mean (sd) & 67.8 (7.6) & 66.3 (8.6) & 69.4
(7.1)\tabularnewline
Individuals affected by breach, mean (sd) & . & . & 39568
(370823)\tabularnewline
PSI-90 composite, mean (sd) & 0.7 (0.2) & 0.8 (0.3) & 0.8
(0.3)\tabularnewline
PSI-4 surgical complication, mean (sd) & 114.9 (18.7) & 113.8 (19.2) &
118.4 (20.6)\tabularnewline
PSI-6 iatrogenic pneumothorax, mean (sd) & 0.4 (0.1) & 0.4 (0.1) & 0.4
(0.1)\tabularnewline
PSI-12 pulmonary embolism, mean (sd) & 4.2 (1.8) & 5.1 (2.5) & 4.9
(2.1)\tabularnewline
PSI-15 accidental laceration, mean (sd) & 1.9 (0.7) & 2.0 (0.8) & 2.0
(0.8)\tabularnewline
\bottomrule
\end{longtable}

Characteristics of the hospital-year observations by breach status are
summarized in Table 3. The timing of breaches varied. Among the breached
hospitals, most of the pre-breach hospital-year observations came from
years 2011-2013 while most post-breach observations came from years
2013-2015. Because of variability in breach event timing, it was
impossible to assign the hospitals that were never breached into a pre-
or post-event category based on time. Therefore, the never-breached
hospital-year observations were pooled into a single control group. This
is a limitation to the comparability of the time-varying characteristics
between the never-breached group and the pre-breach group in Table 3.

The control group and the pre-breach group had similar distributions for
the 30-day AMI mortality rate. The mean 30-day AMI mortality rate for
the pre-breach group was 15.2\%; for the control group it was 15.1\%.

The mean number of beds for the pre-breach group was nearly two times
larger than the control group (469.7 versus 250.4). Among the breached
group, the number of beds was higher in the post-breach group.

The proportion of hospital-year observations that reported being a
\emph{meaningful user} of electronic health records (HITECH) for that
year varied across the control (64.0\%), pre- (41.0\%), and post-breach
(75.8\%) groups. The variation was due to the differences in the
distribution of observations by year for each group. Cross-tabulating
the percentage of meaningful users by group and year showed that the
overall percentage grew from 13\% in 2011 to 94\% by 2014, and that each
group followed a similar time trend.

The proportions of not-for-profit hospitals were similar between the
control group and the breached group. However, the breached group had a
higher proportion of public hospitals, while the control group had a
higher proportion of for-profit hospitals. The breached group was more
likely to be major teaching hospitals. Patient satisfaction measures
were similar between the control group and the breached group, and
satisfaction within the breached group did not vary between the pre- and
post-breach group.

\clearpage

\subsection{Estimates}\label{estimates}

Multivariate regression estimates indicate that a data breach was
associated with a 0.338{[}95\% CI, 0.101-0.576{]} percentage point
increase in the 30-day AMI mortality rate one year after the breach,
0.446{[}95\% CI, 0.164-0.729{]} percentage point increase two years
after the breach, and 0.363{[}95\% CI, 0.0174-0.709{]} percentage point
increase three years after the breach (Table 4). 30-day AMI mortality
rate of breached hospitals did not differ significantly from the
never-breached hospitals in the pre-breach periods.

\begin{longtable}[c]{@{}lll@{}}
\caption{Multivariate model of 30-day AMI mortality rate and hospital
data breach}\tabularnewline
\toprule
Variable & Coefficient Estimate & 95\% CI\tabularnewline
\midrule
\endfirsthead
\toprule
Variable & Coefficient Estimate & 95\% CI\tabularnewline
\midrule
\endhead
Relative time (ref= -1) & &\tabularnewline
-4 & -0.126 & {[}-0.602,0.351{]}\tabularnewline
-3 & 0.0207 & {[}-0.291,0.333{]}\tabularnewline
-2 & -0.0931 & {[}-0.298,0.111{]}\tabularnewline
0 & 0.0846 & {[}-0.101,0.270{]}\tabularnewline
1 & 0.338** & {[}0.101,0.576{]}\tabularnewline
2 & 0.446** & {[}0.164,0.729{]}\tabularnewline
3 & 0.363* & {[}0.0174,0.709{]}\tabularnewline
4 & 0.213 & {[}-0.237,0.664{]}\tabularnewline
operating revenue & 1.54e-10 & {[}-6.67e-11,3.75e-10{]}\tabularnewline
number of beds & 0.000133 & {[}-0.000263,0.000528{]}\tabularnewline
length of stay & 0.0470 & {[}-0.0284,0.122{]}\tabularnewline
bed occupancy rate & 0.178 & {[}-0.304,0.660{]}\tabularnewline
meaningful user (ref=no) & &\tabularnewline
yes & 0.000662 & {[}-0.0653,0.0666{]}\tabularnewline
definitely not recommend hospital & 0.00805 &
{[}-0.00936,0.0255{]}\tabularnewline
year (ref=2011) & &\tabularnewline
2012 & -0.323*** & {[}-0.373,-0.272{]}\tabularnewline
2013 & -0.616*** & {[}-0.692,-0.539{]}\tabularnewline
2014 & -0.916*** & {[}-1.009,-0.823{]}\tabularnewline
2015 & -1.575*** & {[}-1.667,-1.482{]}\tabularnewline
Constant & 15.29*** & {[}14.87,15.72{]}\tabularnewline
N & 11568 &\tabularnewline
N group & 2619 &\tabularnewline
95\% confidence intervals in brackets & &\tabularnewline
* p\textless{}0.05, ** p\textless{}0.01, *** p\textless{}0.001 &
&\tabularnewline
\bottomrule
\end{longtable}

Estimation results that are shown in Table 4 are plotted in Figure 3.
The y-intercept is the expected 30-day AMI mortality rate at one year
before the breach. It is the baseline 30-day AMI mortality if a breach
had not occurred, and for ease of interpretation, we centered it at zero
instead of the grand mean. The plotted points are the expected 30-day
AMI mortality rate at the relative breach time, adjusting for the
baseline rate, yearly time trends, time-invariant hospital effects, and
time-varying hospital characteristics. At 1, 2, 3, years after the
breach, the 30-day AMI mortality rate point estimates are significantly
higher than the baseline.

\begin{figure}[htbp]
\centering
\includegraphics{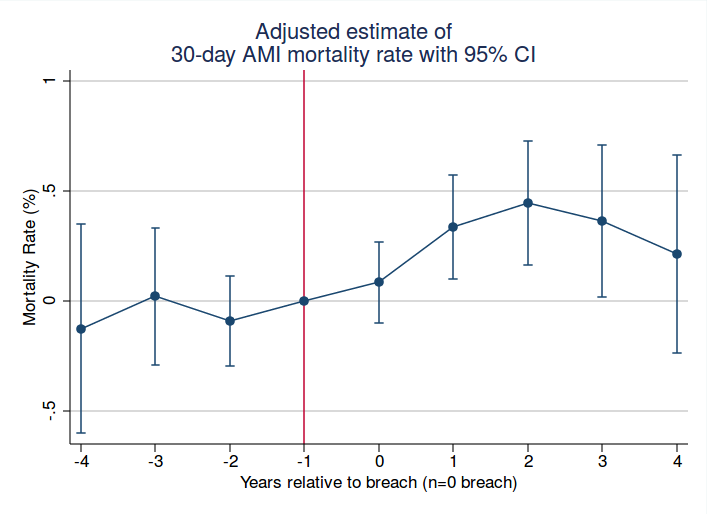}
\caption{Estimated association between breach and 30 day mortality rate
for AMI\label{fig:mrgplot_brcany}}
\end{figure}

We tested whether the association between data breaches and the 30-day
AMI mortality rate was stronger for more serious breaches. Breached
hospitals were categorized into two groups: above or below the median
number of breached individual records. From the reference model, the
pre-breach and post-breach indicators were interacted with an indicator
for the magnitude of breach. The estimation results are plotted in
Figure 4. The association between data breaches and AMI mortality rate
did not differ significantly by the magnitude of the breach.

\begin{figure}[htbp]
\centering
\includegraphics{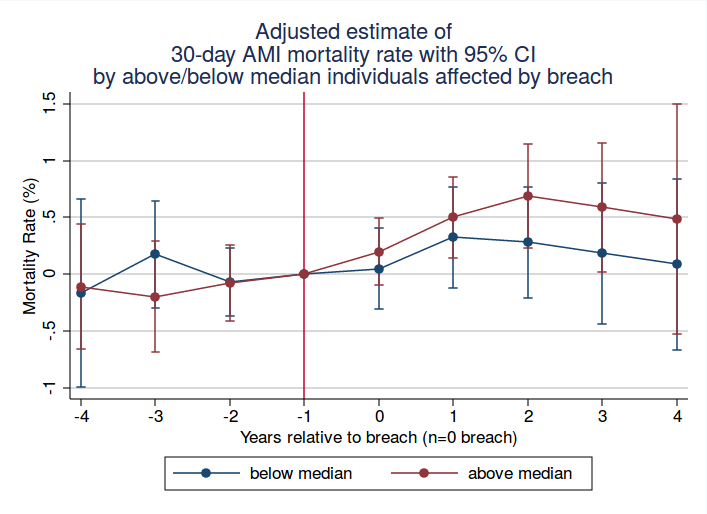}
\caption{Estimated association between breach and 30 day mortality rate
for AMI by severity of breach\label{fig:mrgplot_brcany_indaff}}
\end{figure}

In another test for moderating factors, the pre- post-breach indicators
were interacted with the type of breach. External breaches included
hacking, theft; internal breaches included improper disposal, loss, and
unauthorized disclosure. The relation between breaches and AMI mortality
did not differ significantly by the type of breach (Figure 5).

\begin{figure}[htbp]
\centering
\includegraphics{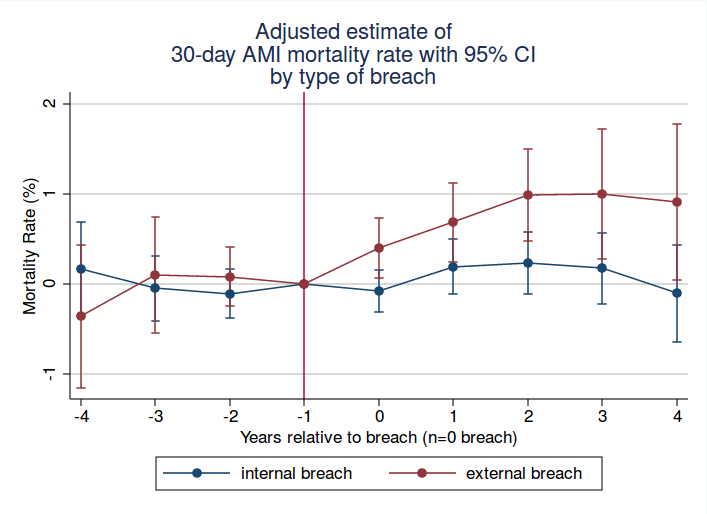}
\caption{Estimated association between breach and 30 day mortality rate
for AMI by type of breach\label{fig:mrgplot_brcany_brctyp}}
\end{figure}

\subsection{Robustness Tests}\label{robustness-tests}

Care quality problems that involve provider error or negligence may be
correlated with both data breaches and patient mortality. Patient safety
indicators were added to the reference model to control for care
quality. Patient safety indicators were only available for a subset of
observations. Patient safety indicators PSI-4, PSI-6, PSI-12, PSI-15
were available from 2012 (7,096 observations). PSI-90 composite index
was available from 2013 (5,227 observations). The following alternative
models were estimated: (1) reference model estimated with the subset,
(2) model estimated with PSI-90 composite index, and (3) model estimated
with PSI-4, PSI-6, PSI-12, and PSI-15. Including the patient safety
indicators did not change the results from the reference model. We
conclude that our model findings are robust to hospital differences in
overall care quality.

\section{Discussion}\label{discussion}

We find that hospital data breaches were associated with higher 30-day
AMI mortality rates in the years following the breach. Figure 1 shows
that improvements in AMI treatment have resulted in the 30-day AMI
mortality rate decreasing about 0.4 percentage points annually since
2011. The .34 to.45 percentage point increase in 30-day AMI mortality
rate after a breach was comparable to undoing a year's worth of
improvement in mortality rate. The national estimate for the number of
hospital discharges for AMI has fluctuated around 556,000 discharges
annually between 2005 and 2014.{[}31{]} On average, a data breach at a
non-federal acute-care inpatient hospital was associated with an
additional 34 to 45 deaths per 1000 AMI discharges per year.

Changes in HIT and patient care processes in response to a data breach
introduce usability challenges and unintended side effects that
frustrate clinicians and disrupt patient care.{[}32{]} Frustrated
clinicians bypassing the new system and process with ad hoc workarounds
creates new opportunities for errors.{[}33,34{]} Enhanced security
measures in response to a data breach are likely to worsen the usability
of the HIT system, which would not only diminish the effectiveness of
its intended function but also spawn new errors that worsen the quality
of care provided to patients.

We observed that the elevated mortality rates persisted for three years
subsequent to a breach. We believe that the impact of a breach is likely
shorter than three years as hospitals tend to implement new procedures,
processes, and technologies in the first year following a breach. Thus
we expect the impact of the breach would dissipate over time -- possibly
less than two years. We note that the data generating process for
mortality rates may contribute to the three-year observed impact.
Hospital Compare Data measure 30-day mortality as a 36-month moving
average, which would smooth the observed response to a breach. The
higher mortality rate at three years after the breach suggests this
smoothing may be extending the observed time-impact.

More recently, the emergence of hospital ransomware attacks have
disrupted hospital services and there are growing fears of attacks on
the care delivery system itself.{[}35{]} Ransomware attacks are
considered to be more disruptive to hospital operations than the
breaches considered in this study. The data breaches studied in our
analysis come from 2011-2015 when such ransomware or infrastructure
attacks were rare. If disruption to information technology used by
providers is driving the breach effect, the findings from our study
suggest that ransomware attacks may have an even stronger negative
impact on patients than the breaches studied in this paper.

Time varying care quality problems are potential confounders to
estimating the breach impact on patient outcomes. Controlling for
patient safety indicators attempted to address this concern. The breach
impact estimates were similar between the models with and without the
patient safety indicators. The findings suggest that patient safety
indicators were not confounding factors, but raises new concerns whether
these indicators were effective controls for care quality problems.

Health data breaches have significant consequences for patients,
providers, and payers, which could be framed as a quality of care
problem. Protecting health information should be an important
responsibility of all parties in the healthcare industry. Our results
indicate that breaches and the subsequent hospital reaction may
adversely impact care quality. We suggest that breached hospitals should
carefully consider subsequent security initiatives to reduce the
potential impact of new processes, procedures, and technologies on care
quality. The healthcare community must work together to jointly address
the need to protect patient data and improve patient outcomes.

\section*{References}\label{references}
\addcontentsline{toc}{section}{References}

\hyperdef{}{ref-Humer2014}{\label{ref-Humer2014}}
{[}1{]} Humer C, Finkle J. Your medical record is worth more to hackers
than your credit card 2014.

\hyperdef{}{ref-Beek2016}{\label{ref-Beek2016}}
{[}2{]} Beek C, McFarland C, Samani R. Health Warning: Cyberattacks are
targeting the health care industry. Intel Security; 2016.

\hyperdef{}{ref-hhs2016f}{\label{ref-hhs2016f}}
{[}3{]} U.S. Department of Health \& Human Services. HIPAA Compliance
and Enforcement 2016.

\hyperdef{}{ref-Wong2016}{\label{ref-Wong2016}}
{[}4{]} Wong J. Los Angeles hospital returns to faxes and paper charts
after cyberattack 2016.

\hyperdef{}{ref-Cox2016}{\label{ref-Cox2016}}
{[}5{]} Cox J, Turner K, Zapotosky M. Virus infects MedStar Health
system's computers, forcing an online shutdown 2016.

\hyperdef{}{ref-handoff2016}{\label{ref-handoff2016}}
{[}6{]} Agency for Healthcare Research and Quality. Handoffs and
Signouts 2016.

\hyperdef{}{ref-Pham2012}{\label{ref-Pham2012}}
{[}7{]} Pham JC, Aswani MS, Rosen M, Lee H, Huddle M, Weeks K, et al.
Reducing Medical Errors and Adverse Events. Annual Review of Medicine
2012;63:447--63.
\href{http://doi.org/10.1146/annurev-med-061410-121352}{doi:10.1146/annurev-med-061410-121352}.

\hyperdef{}{ref-Borowitz2008}{\label{ref-Borowitz2008}}
{[}8{]} Borowitz SM, Waggoner-Fountain LA, Bass EJ, DeVoge JM. Resident
Sign-Out: A Precarious Exchange of Critical Information in a Fast-Paced
World. Agency for Healthcare Research; Quality (US); 2008.

\hyperdef{}{ref-Graham2013}{\label{ref-Graham2013}}
{[}9{]} Graham KL, Marcantonio ER, Huang GC, Yang J, Davis RB, Smith CC.
Effect of a Systems Intervention on the Quality and Safety of Patient
Handoffs in an Internal Medicine Residency Program. Journal of General
Internal Medicine 2013;28:986--93.
\href{http://doi.org/10.1007/s11606-013-2391-7}{doi:10.1007/s11606-013-2391-7}.

\hyperdef{}{ref-PonemonInstitute2016}{\label{ref-PonemonInstitute2016}}
{[}10{]} Ponemon Institute. 2016 Cost of Data Breach Study: United
States. Ponemon Institute; 2016.

\hyperdef{}{ref-Romanosky2016}{\label{ref-Romanosky2016}}
{[}11{]} Romanosky S, Anderson J, Crespi M, Gordon L, Graves J,
Greisiger M, et al. Examining the Costs and Causes of Cyber Incidents.
Journal of Cybersecurity 2016:1--15.
\href{http://doi.org/10.1093/cybsec/tyw001}{doi:10.1093/cybsec/tyw001}.

\hyperdef{}{ref-hhs2016d}{\label{ref-hhs2016d}}
{[}12{]} U.S. Department of Health \& Human Services. Resolution
Agreements 2016.

\hyperdef{}{ref-hhsocr2016a}{\label{ref-hhsocr2016a}}
{[}13{]} U.S. Department of Health \& Human Services. Breach
Notification Rule 2016.

\hyperdef{}{ref-hhsocr2016}{\label{ref-hhsocr2016}}
{[}14{]} U.S. Department of Health \& Human Services. Breach Report
2016.

\hyperdef{}{ref-hhsocr2016c}{\label{ref-hhsocr2016c}}
{[}15{]} U.S. Department of Health \& Human Services. Methods for
De-identification of PHI 2016.

\hyperdef{}{ref-prc2016}{\label{ref-prc2016}}
{[}16{]} Privacy Rights Clearinghouse. Privacy Rights Clearinghouse
Chronology of Data Breaches 2016.

\hyperdef{}{ref-sb1386}{\label{ref-sb1386}}
{[}17{]} California Legislature. S.B. 1386, 2001-02 Leg., Reg. Sess.
(Cal. 2002), codified at CAL. CIV. CODE § 1798.29, 1798.80-.84 (2009)
2002.

\hyperdef{}{ref-ncls2016}{\label{ref-ncls2016}}
{[}18{]} National Conference of State Legislatures. Security Breach
Notification Laws 2016.

\hyperdef{}{ref-CMS2016}{\label{ref-CMS2016}}
{[}19{]} Centers for Medicare \& Medicaid Services. Healthcare Cost
Report Information System 2016.

\hyperdef{}{ref-CMSCompare2016}{\label{ref-CMSCompare2016}}
{[}20{]} Centers for Medicare \& Medicaid Services. Medicare Hospital
Compare 2016.

\hyperdef{}{ref-Jacobson1993}{\label{ref-Jacobson1993}}
{[}21{]} Jacobson LS, Lalonde RJ, Sullivan DG. American Economic
Association Earnings Losses of Displaced Workers Earnings Losses of
Displaced Workers. The American Economic Review 1993;83:685--709.

\hyperdef{}{ref-CMS2016a}{\label{ref-CMS2016a}}
{[}22{]} Centers for Medicare \& Medicaid Services. Acute myocardial
infarction (AMI): hospital 30-day, all-cause, risk-standardized
mortality rate (RSMR) following AMI hospitalization. \textbar{} National
Quality Measures Clearinghouse 2016.

\hyperdef{}{ref-McCullough2013}{\label{ref-McCullough2013}}
{[}23{]} McCullough J, Parente S, Town R. Health Information Technology
and Patient Outcomes: The Role of Organizational and Informational
Complementarities 2013.
\href{http://doi.org/10.3386/w18684}{doi:10.3386/w18684}.

\hyperdef{}{ref-Hansen2011}{\label{ref-Hansen2011}}
{[}24{]} Hansen LO, Williams MV, Singer SJ. Perceptions of hospital
safety climate and incidence of readmission. Health Services Research
2011;46:596--616.
\href{http://doi.org/10.1111/j.1475-6773.2010.01204.x}{doi:10.1111/j.1475-6773.2010.01204.x}.

\hyperdef{}{ref-kwon2015}{\label{ref-kwon2015}}
{[}25{]} Kwon J, Johnson ME. The Market Effect of Healthcare Security:
Do Patients Care about Data Breaches? 2015.

\hyperdef{}{ref-Zohar1980}{\label{ref-Zohar1980}}
{[}26{]} Zohar D. Safety climate in industrial organizations:
theoretical and applied implications. The Journal of Applied Psychology
1980;65:96 102.

\hyperdef{}{ref-Singer2009}{\label{ref-Singer2009}}
{[}27{]} Singer SJ, Falwell A, Gaba DM, Meterko M, Rosen A, Hartmann CW,
et al. Identifying organizational cultures that promote patient safety.
Health Care Management Review 2009;34:300--11.
\href{http://doi.org/10.1097/HMR.0b013e3181afc10c}{doi:10.1097/HMR.0b013e3181afc10c}.

\hyperdef{}{ref-CMS2015}{\label{ref-CMS2015}}
{[}28{]} Centers for Medicare \& Medicaid Services. Hospital Consumer
Assessment of Healthcare Providers and Systems (HCAHPS) Survey 2015.

\hyperdef{}{ref-Friedman2009}{\label{ref-Friedman2009}}
{[}29{]} Friedman B, Encinosa W, Jiang HJ, Mutter R. Do Patient Safety
Events Increase Readmissions? Medical Care 2009;47:583--90.
\href{http://doi.org/10.1097/MLR.0b013e31819434da}{doi:10.1097/MLR.0b013e31819434da}.

\hyperdef{}{ref-Wang2016}{\label{ref-Wang2016}}
{[}30{]} Wang Y, Eldridge N, Metersky ML, Sonnenfeld N, Fine JM, Pandol
MM, et al. Association Between Hospital Performance on Patient Safety
and. Journal of the American Heart Association 2016;5:1--14.
\href{http://doi.org/10.1161/JAHA.116.003731}{doi:10.1161/JAHA.116.003731}.

\hyperdef{}{ref-AgencyforHealthcareResearchandQuality2016a}{\label{ref-AgencyforHealthcareResearchandQuality2016a}}
{[}31{]} Agency for Healthcare Research and Quality. HCUP Projections
Acute Myocardial Infarction and Acute Stroke 2005 to 2016. 2016.

\hyperdef{}{ref-Koppel2016}{\label{ref-Koppel2016}}
{[}32{]} Koppel R. Great Promises of Healthcare Information Technology
Deliver Less. In:, Springer International Publishing; 2016, pp. 101--25.
\href{http://doi.org/10.1007/978-3-319-20765-0_6}{doi:10.1007/978-3-319-20765-0\_6}.

\hyperdef{}{ref-Campbell2006}{\label{ref-Campbell2006}}
{[}33{]} Campbell EM, Sittig DF, Ash JS, Guappone KP, Dykstra RH. Types
of unintended consequences related to computerized provider order entry.
Journal of the American Medical Informatics Association : JAMIA
2006;13:547--56.
\href{http://doi.org/10.1197/jamia.M2042}{doi:10.1197/jamia.M2042}.

\hyperdef{}{ref-Han2005}{\label{ref-Han2005}}
{[}34{]} Han YY, Carcillo JA, Venkataraman ST, Clark RSB, Watson RS,
Nguyen TC, et al. Unexpected Increased Mortality After Implementation of
a Commercially Sold Computerized Physician Order Entry System.
PEDIATRICS 2005;116:1506--12.
\href{http://doi.org/10.1542/peds.2005-1287}{doi:10.1542/peds.2005-1287}.

\hyperdef{}{ref-Green2016}{\label{ref-Green2016}}
{[}35{]} Green M. Hospitals are hit with 88\% of all ransomware attacks
2016.

\end{document}